\documentclass[9pt]{sig-alternate} 
\usepackage[usenames,dvipsnames]{color}
\usepackage[utf8]{inputenc} 

\usepackage{graphicx} 

\usepackage{paralist} 
\usepackage{listings}

\usepackage{mathtools}
\usepackage{multirow}
\usepackage{makecell}
\usepackage{flushend}
\DeclarePairedDelimiter\abs{\lvert}{\rvert}%
\DeclarePairedDelimiter\norm{\lVert}{\rVert}%
\usepackage{booktabs}

\makeatletter
\let\oldabs\abs
\def\abs{\@ifstar{\oldabs}{\oldabs*}}
\let\oldnorm\norm
\def\norm{\@ifstar{\oldnorm}{\oldnorm*}}

\renewcommand{\paragraph}[1]{\vskip0.1in\noindent {\bf{#1}}}

\newcommand{\etal}{\textit{et~al}.}

\begin{document}

\title{Inferring Person-to-person Proximity Using WiFi Signals}

\numberofauthors{4}
\author{
\alignauthor
Piotr Sapiezynski\\
	\affaddr{Technical University of Denmark}\\
	\email{pisa@dtu.dk}
\alignauthor
Arkadiusz Stopczynski\\
	\affaddr{Technical University of Denmark}\\
	\affaddr{MIT Media Lab}\\
	\email{arks@dtu.dk}
\alignauthor
David Kofoed Wind\\
	\affaddr{Technical University of Denmark}\\
	\email{dawi@dtu.dk}
\and
\alignauthor
Jure Leskovec\\
	\affaddr{Stanford University}\\
	\email{jure@cs.stanford.edu}
\alignauthor
Sune Lehmann\\
	\affaddr{Technical University of Denmark,}\\
	\affaddr{Niels Bohr Institute}\\
	\email{sljo@dtu.dk}
}

\maketitle

\begin{abstract}
Today's societies are enveloped in an ever-growing telecommunication infrastructure.
This infrastructure offers important opportunities for sensing and recording a multitude of human behaviors.
Human mobility patterns are a prominent example of such a behavior which has been studied based on cell phone towers, Bluetooth beacons, and WiFi networks as proxies for location.
However, while mobility is an important aspect of human behavior, understanding complex social systems requires studying not only the movement of individuals, but also their interactions.
Sensing social interactions on a large scale is a technical challenge and many commonly used approaches---including RFID badges or Bluetooth scanning---offer only limited scalability.
Here we show that it is possible, in a scalable and robust way, to accurately infer person-to-person physical proximity from the lists of WiFi access points measured by smartphones carried by the two individuals.
Based on a longitudinal dataset of approximately 800 participants with ground-truth interactions collected over a year, we show that our model performs better than the current state-of-the-art.
Our results demonstrate the value of WiFi signals in social sensing as well as potential threats to privacy that they imply.
\end{abstract}



%


\section{Introduction}

We are surrounded by an ever-increasing number of telecommunication infrastructures, such as mobile phone networks, WiFi access points, or Bluetooth beacons.
In addition to their intended function of providing connectivity, these infrastructures offer an unprecedented opportunity for sensing, modeling, and subsequent analyzing of a wide range of human behaviors~\cite{lazer2009life}.
Here we show how our interactions with other people can be inferred in a reliable and scalable way, using signals from WiFi access points.

Being able to infer person-to-person proximity events with high spatio-temporal resolution enables modeling of phenomena such as spreading of diseases and information~\cite{isella2011s}, formation of social ties~\cite{eagle2009inferring}, as well as group dynamics~\cite{sekara2015fundamental}. 
Commercial applications vary from distributed ad hoc networking~\cite{Li2000} to romantic matchmaking~\cite{happn}.

Despite the importance of understanding networks of close proximity interactions, there is a scarcity of scalable and efficient ways to obtain data for large populations.
This is due to the fact that technology has only recently developed to the point, where collection of such high resolution data has become technologically feasible.
The data sources used for investigating mobility of individuals, such as call detail records (CDRs) from mobile operators~\cite{gonzalez2008understanding}, are too coarse in terms of temporal and spatial resolution to allow inference of person-to-person proximity.
On the other hand, the current state-of-the-art methods for measurement of physical proximity require using specialized hardware (\textit{e.g.}, sociometric badges)~\cite{olguin2009sensible, salathe2010high} or smartphones sensing each other through Bluetooth~\cite{eagle2006reality, aharony2011social, 10.1371/journal.pone.0095978}.
Specialized hardware adds cost and complexity to experimental deployments, effectively limiting their scale.
Bluetooth scanning realized on participants' mobile phones increases power consumption~\cite{6200281}---limiting temporal resolution that can be achieved---and requires the devices to be in Bluetooth \emph{discoverable mode}.
This requirement raises privacy~\cite{btprivacy} and security concerns~\cite{scarfone2008guide}. 
When a phone is in discoverable mode the location of its owner can be tracked by third parties, a fact commonly used by researchers~\cite{Larsen2013roskilde,o2006instrumenting}, and advertisers~\cite{guardianBluetooth}.
Moreover, whenever a phone is discoverable, a malicious actor can attempt to pair to it in order to steal contact lists or content of messages.
For these reasons phone manufacturers make it difficult (or impossible) for a handset to remain discoverable indefinitely. 
iOS and Android 6.0+ devices disable discoverability whenever the user exits the Bluetooth settings screen.
Older Android devices let the user set the discoverability timeout to, at maximum, five minutes.
In our study we relied on the fact that in Android versions 4.1 - 6.0 it is still possible to set unlimited discoverability timeout programmatically, but this might change at any point in the future.
Apart from the privacy and security issues of using Bluetooth for sensing, another shortcoming is that Bluetooth data lacks location context.
When co-presence of individuals is inferred through devices sensing each other, an additional step is usually required to estimate the location of the meeting, for example by comparing Bluetooth scans with GPS measurements~\cite{sekara2015fundamental}, by using fixed infrastructure of RFID transmitters~\cite{stehle2011high}, or Bluetooth beacons~\cite{Larsen2013roskilde}.
In the light of these problems, it is clear that alternative methods for tracking person-to-person interactions are needed.
There have been attempts at exploiting WiFi signals for social sensing (\textit{e.g.},~\cite{meunier2004peer,krumm2004nearme,McNett2005,kjaergaard2012challenges} further described in the related work section) but their general applicability is unclear.
The previous methods relied on a single feature for comparing list of detected WiFi devices, they were only trained and tested in controlled environments, and they lack verification on longer timescales.


\paragraph{Present work.}
Here we study the problem of inferring physical proximity between pairs of individuals from a list of WiFi signals sensed by their phones.
We use a longitudinal dataset containing WiFi and Bluetooth scan results from hundreds of participants, collected over a year as part of the Copenhagen Network Study~\cite{10.1371/journal.pone.0095978}.
Using Bluetooth as ground-truth for physical proximity, we train a model for comparing the results of WiFi scans from two devices to determine whether two individuals were in close physical proximity.
We employ a number of interpretable metrics to compare the lists of visible WiFi access points, such as Jaccard similarity or correlation of received signal strengths.
Apart from comparing the lists directly, we can derive context from just the number of routers seen in the lists: more populated areas tend to have more routers available.
Furthermore, we exploit the characteristics of interaction dynamics, for example that people are more likely to meet during work hours, or on a Friday afternoon than on a Sunday night.
Importantly, our algorithm for using WiFi signals to infer proximity does not rely on positioning the routers in physical space. 
Co-location is not inferred by thresholding the distance between the estimated location of two individuals.
Instead, their WiFi environments are compared and then we estimate the similarity directly.
As a final step, we are able to combine these insights using machine learning models to achieve the area under receiving operator curve (AUC ROC) scores of up to $0.89$ in the proximity inference task.
We show that our model works in a range of environments, does not depend on particular access points, and its performance does not deteriorate over time.
Our experiments demonstrate that we are able to track close-proximity interactions over time and in different social and spatio-temporal contexts.
Overall, our approach performs better than previously suggested solutions.

\paragraph{Contribution.}
We present a novel approach for tracking close-proximity person-to-person interactions based on existing infrastructure of WiFi networks and off-the-shelf consumer smartphones and compare its performance against existing methods.


\section{Experimental design}
The dataset used in this work was collected as part of the Copenhagen Networks Study~\cite{10.1371/journal.pone.0095978}.
It covers mobility and interaction records of approximately 1\,000 students at Technical University of Denmark, over a two year period.
Each student was equipped with a LGE Nexus 4 Android smartphone as a data collecting device.
On each phone, an application based on the Funf Open Sensing framework~\cite{aharony2011social} gathered readings from multiple sensors including:
\begin{itemize}
\item Bluetooth scans (every 5 minutes): each scan contains a list of discoverable devices,\footnote{smartphones in the study were specifically configured to be in Bluetooth discoverable mode} their unique identifiers, user defined names, and received signal strength (RSSI). Because we know which anonymized participant identifier corresponds to which Bluetooth unique identifier, we can monitor proximity between the participants.
\item WiFi scans (every 5 minutes): each scan contains a list of WiFi access points (both traditional routers and mobile hotspots), their unique identifiers (BSSIDs or MAC addresses), network names they transmit (SSIDs), and RSSI.
\end{itemize}

The collector app additionally collected the data requested by other applications on the phone.
Therefore, the temporal resolution of the data for some of the users can be even higher than one sample every 5 minutes.

All data in the Copenhagen Networks Study was collected with the participants' informed consent, with an emphasis on ensuring awareness of the complexity and sensitivity of the collected data~\cite{StopczynskiPPLL14}.
The study setup, including security, privacy, and informed consent has been approved by Danish Data Protection Agency.
Further details of the study can be found in Ref.~\cite{10.1371/journal.pone.0095978}.




\section{Methods}
\label{section:methods}

In brief, our task is to compare the lists of WiFi routers seen by users $A$ and $B$ approximately at the same time (with at most $\Delta t=300$ seconds difference) and determine whether the two users were in close physical proximity.
We use Bluetooth data as ground truth for physical proximity to train and verify our models.

\begin{table}[t]
  \centering
  \begin{tabular}{p{40mm}|c|c}
  	\toprule
  	& training & test \\
  	\toprule
  	total observations & 0.5M & 115.5M \\
  	\% positive & 31\% & 31\% \\
  	unique users & 812 & 820 \\
  	median number of access points per observation & 7.0 & 7.0 \\
    mean number of access points per observation & 11.3 & 11.3 \\
  	\bottomrule
  \end{tabular}
  \caption{Summary statistics of the dataset used to infer proximity events.}
  \label{tab:summary_stats_proximity}
\end{table}

\subsection{Data preparation}
\paragraph{WiFi. }
We found that in our dataset there are multiple WiFi routers that share the same MAC address, a phenomenon which might confound our task. 
We use a simple heuristic to remove these ``ambiguous'' routers since finding the optimal way of identifying them would warrant a publication on its own. 
Here we rely on the network name they broadcast. 
Because the routers at the DTU campus broadcast up to four network names (SSID) per MAC address, we remove the scans of routers which broadcast five or more network names throughout the observation.
We found 3950 offending MAC addresses, which corresponds to only 0.04\% of all unique MAC addresses in the data.
However, scans of these routers constitute 1.4\% of all scan results.

Next, we identify one home router for each participant per month. 
We employ the following heuristic for each participant:
\begin{enumerate}
\item Bin the time information of WiFi scan history. The size of the bin does not influence the results significantly, here we use 10 minutes.
\item Sort the list of routers by the number of timebins in which they appear, in descending order.
\item The router that appears in the biggest number of timebins is assumed to be the home router.
\end{enumerate}
The details of the procedure are described in Ref.~\cite{10.1371/journal.pone.0130824}.

\paragraph{Bluetooth. }
Due to the imperfect firmware and software running on the phones, Bluetooth data is not always available---not all users are scanning and discoverable at all times.
This can introduce a situation in which two persons are proximate, but Bluetooth does not capture that event. 
We divide the dataset into one hour subsets and select only the WiFi and Bluetooth data from people who were seen and who saw at least one other person through Bluetooth. 
This strict approach makes the task more difficult, as it removes long periods where individuals are alone, for example night-time samples of students who do not live with other participants.

\paragraph{Negative samples. }
To train our model we also need to provide negative examples.
For dyads in this category we choose potential interactions between two people who did not see each other on Bluetooth, but whose lists of scan results share at least one overlapping router.
Compared to selecting negative samples by randomly sampling dyads this definition brings the task closer to a real-life scenario of discovering very close physical proximity (up to approximately 10 meters).
As a result, the dataset has 31\% positive and 69\% negative samples.

\subsection{Dataset statistics}
Table~\ref{tab:summary_stats_proximity} shows the details about the dataset. 
Through a year of data we found 116M potential interactions. We randomly select 0.5M of them to train the models.

\begin{figure}[t!]
\center
\includegraphics[width=1\linewidth]{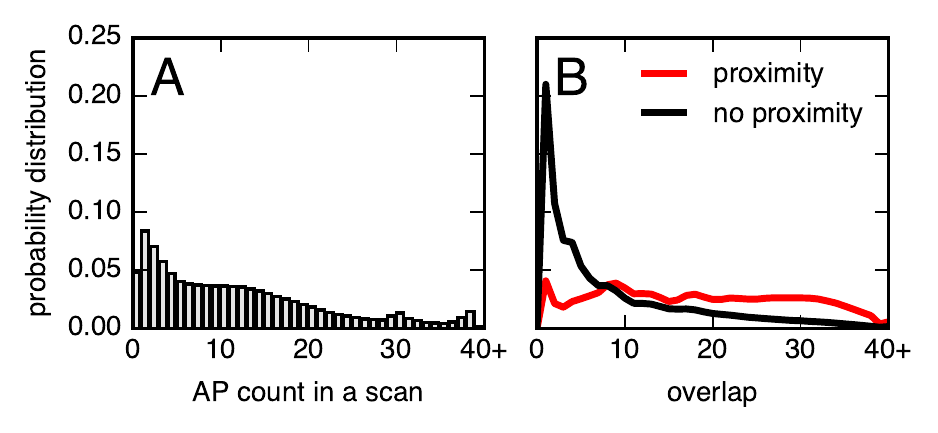}

\caption{a. More than 95\% of scans report at least one access point, and 12 APs on average. b. People in Bluetooth proximity scan more overlapping routers than those who are not proximate.}
\label{fig:count_overlap}
\end{figure}

We note that in our dataset people are near to access points more than 95\% of time, and the average count of routers in a single scan is 12, see Figure~\ref{fig:count_overlap}A.
We also observe that in 99\% of cases of Bluetooth sightings the corresponding WiFi scans overlap by at least one access point. This indicates that there is a potential in using WiFi scan results to infer the co-presence with high recall.
Conversely, in more than 31\% of cases where there is at least one overlapping access point, the two devices are also close according to Bluetooth. 
This indicates that WiFi signals can be applied to the task resulting in a high precision solution.
In general, pairs of people who are in Bluetooth proximity scan more routers in common than those who are not, see Figure~\ref{fig:count_overlap}B.
The majority (53\%) of meetings happen during working hours (from 8am to 7pm) on campus.

\subsection{Methods of comparison}
\begin{figure}[t!]
\center
\includegraphics[width=1\linewidth]{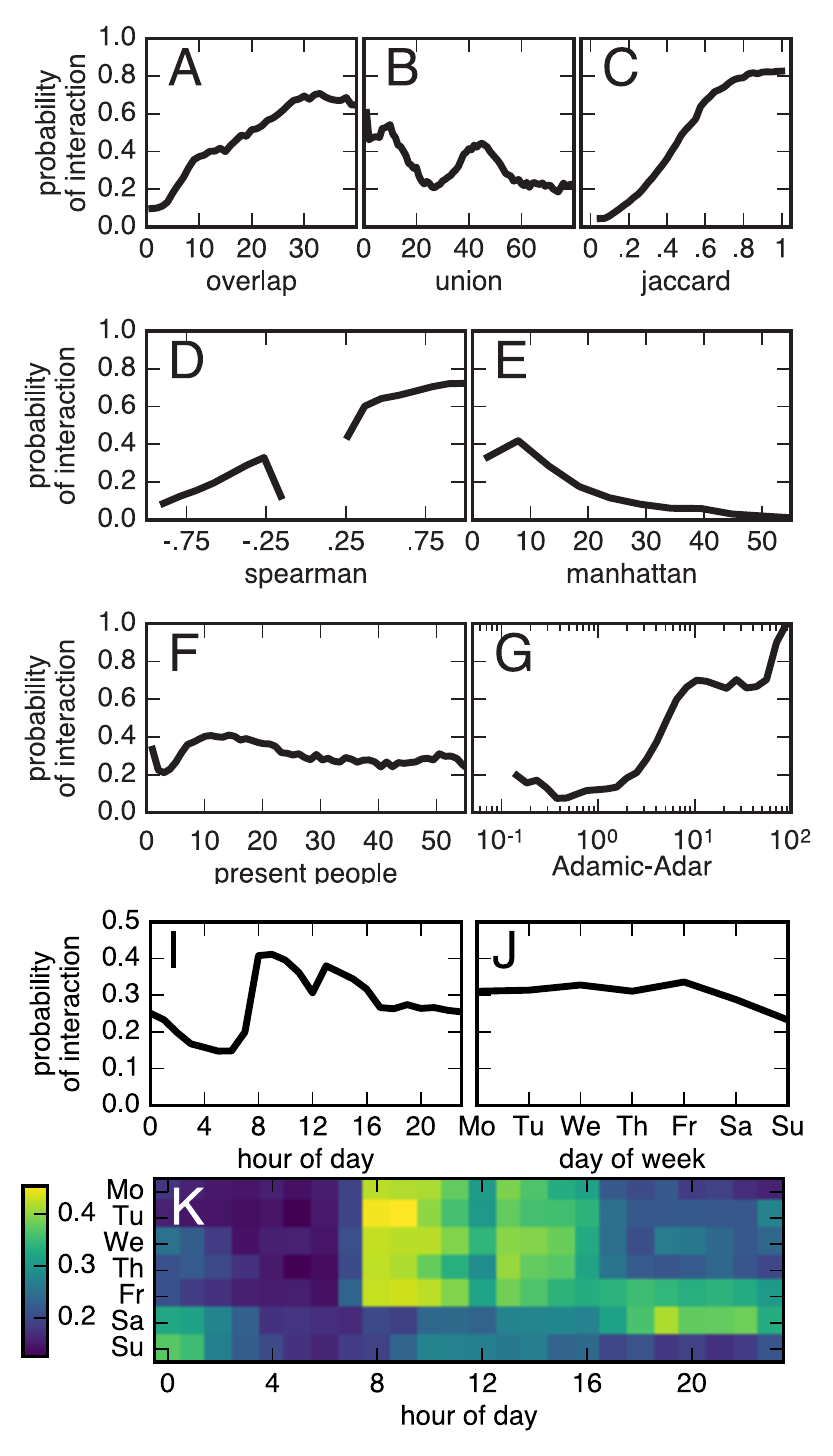}

\caption{The larger the number of common routers two phones see, the higher the probability of close proximity.
At the same time, the more routers they see in total, the lower the probability of an interaction --- densely populated areas have more routers and more people who are not necessarily interacting.
Jaccard similarity allows us to recognize interactions regardless of the number of visible access points.}
\label{fig:bt_wifi_features}
\end{figure}

We use a number of metrics to compare two lists of WiFi scan results and use these metrics as features in a supervised machine learning approach.
We divide the features into the following categories: availability of access points, received signal strength, presence + RSSI, timing, popularity, and location.
Table~\ref{tab:features} lists the features we apply, and Figure~\ref{fig:bt_wifi_features} shows how the probability of an interaction changes as a function of each feature's value.
In this section we describe each feature in detail. 
Citations refer to the first articles using the features for the purpose of person to person contact detection. 

\begin{table}[t]
  \centering
  \begin{tabular}{p{30mm}p{40mm}}
  	\toprule
  	\textbf{category} & \textbf{features} \\
  	\toprule
  	\textbf{AP presence} & overlap, non-overlap, union, jaccard \\ \midrule
  	\textbf{RSSI} & spearman, pearson, manhattan, euclidean \\ \midrule
  	\textbf{AP presence + RSSI} & top AP, top AP$\pm6dB$ \\  \midrule
  	\textbf{timing} & hour of week \\  \midrule
  	\textbf{popularity} & min popularity, max popularity, Adamic-Adar \\ \midrule
    \textbf{location} & at home, at DTU \\ \bottomrule

  \end{tabular}
  \caption{Features used to infer close-proximity interactions.}
  \label{tab:features}
\end{table}

\paragraph{Availability of access points (AP presence).}
First, we compare the list of routers seen by the two phones, regardless of their received signal strengths. 
We introduce the following measures: 
\textbf{overlap}: the raw count of overlapping routers~\cite{krumm2004nearme};
\textbf{union}: size of the union of the two lists;
\textbf{jaccard}: ratio between the size of the intersection and the size of the union of the two lists~\cite{kjaergaard2012challenges}.
\textbf{non-overlap}: the raw count of non-overlapping routers (size of union minus size of overlap)~\cite{krumm2004nearme};
Figure~\ref{fig:bt_wifi_features}A-C presents the interplay between the values of the three parameters and the probability of an interaction.
Intuitively, the greater the number of common routers two phones see in a scan, the higher the probability of them being in close proximity.
Perhaps surprisingly, this probability also depends on the size of the union: the larger the union of the two lists the lower the probability of an interaction.
This can be explained by the fact that the number of available access points is positively correlated with the population density~\cite{10.1371/journal.pone.0130824}.
Hence, popular places are likely to attract people who do not necessarily interact with one another.
Conversely, two people in a relatively unpopular location are more likely to be there together.
The visible dip in the union plot, corresponding to lower probability of meeting with around 30 routers present, might correspond to a particular location where many non-interactions happen (for example a dining hall).
Nevertheless, we expect that, in general, the probability of interaction is negatively correlated with the size of union.
Using Jaccard similarity between the two lists allows to recognize interactions regardless of the number of visible access points.



\paragraph{Received Signal Strength Indicator (RSSI).}
Next, we focus on comparing the received signal strength of the overlapping routers.
While received signal strength (RSSI) is not generally a reliable proxy for distance~\cite{sapiezynski2015opportunities}, two co-located people can be expected to have similar RSSI readings for the overlapping routers.
We investigate the \textbf{spearman} and \textbf{pearson} correlation coefficients of received signal strengths of the overlapping routers.
For brevity we only present the results for the spearman metric Figure~\ref{fig:bt_wifi_features}D --- the values of the two metrics are highly correlated (Spearman's $\rho=0.89$, $p_{val} < 0.001$).
Note that because there are instances where the correlation is undefined (\emph{not a number}) or not statistically significant (with $p_{val}>0.05$), we replace such values of the coefficients with the mean values of valid correlations (see section~\ref{section:imputing} for details of the imputation).
This implies that there are no examples of small correlations (which, given only a few values to compare, are not statistically significant) and there is a dip in probability of interactions corresponding to the mean value of correlation coefficients.

Furthermore, we also calculate the difference between RSSI of overlapping routers by measuring the $\ell_1$ and $\ell_2$ distances and dividing the results by the number of overlapping routers.
For simplicity we call these features \textbf{manhattan} and \textbf{euclidean} and define them in Equations~\ref{eq:manhattan}~\cite{meunier2004peer} and~\ref{eq:euclidean}~\cite{kjaergaard2012challenges} respectively.

\begin{equation}
m=\frac{\sum\limits_{i} \abs{RSSI_{A,i}-RSSI_{B,i}}}{N}
\label{eq:manhattan}
\end{equation}

\begin{equation}
e=\frac{\sqrt{\sum\limits_{i} (RSSI_{A,i}-RSSI_{B,i})^2}}{N}
\label{eq:euclidean}
\end{equation}
where $RSSI_{A,i}$ is the received signal strength or access point $i$ as measured by user $A$, and $N$ is the total number of overlapping routers.
Figure~\ref{fig:bt_wifi_features}E shows that with growing distance, the probability of an interaction falls.

\paragraph{AP presence + RSSI.}
It has been previously shown that a good heuristic for determining whether a user is in the same location during two measurements is to verify whether they measure a common strongest router~\cite{IMM2013-06663}.
Here, we verify whether this approach can be used for inferring co-location: if two users measure the same router as the strongest one, we assume they are in close proximity.
We investigate the strict case, \textbf{top AP}.
Additionally, we allow for some variability in the measured strength: feature \textbf{top AP$\pm6dB$} assumes a positive value if there is at least one overlapping access point in the lists of routers of $A$ and $B$ within $6dB$ from the top router.



\paragraph{Popularity.}
Additionally, we inspect how many different participants of the study scanned the overlapping routers within five minutes of the meeting---intuitively if only a few persons were in a given location they were more likely to be there together, rather than by chance.
We find the least and the most popular among the overlapping routers and report \textbf{min\_popularity} and \textbf{max\_popularity}.
As we show in Figure~\ref{fig:bt_wifi_features}F, this intuition is not entirely confirmed by the data.
The correlation between the number of individuals present and the probability that any two of them are interacting is low (Spearman's $\rho=0.15$, $p_{val}<0.001$).
Note that popularity and the size of union are correlated (Spearman's $\rho=0.48$, $p_{val} < 0.001$) --- more routers are located in popular places, so the more routers there are around, the more people see each of them.
However, to achieve a good estimation of popularity, we need data from the entire population, while the number of routers around can be obtained just from data of just the two individuals.
Additionally, we use a score inspired by a measure introduced by Adamic and Adar~\cite{adamic2003friends}, defined as: 

\begin{equation}
aa(u_1, u_2)=\sum\limits_{i}{\frac{1}{\log(popularity(AP_i))}}.
\label{eq:adamic_adar}
\end{equation}

Here, each overlapping router is weighted more the fewer people scanned it. In this case, the higher the value, the higher the probability of a meeting between two people.

\paragraph{Timing.}
In contrast to the other features we described, timing does not rely on comparing the list of scan results.
Instead, we use the timestamp of each potential meeting to exploit the temporal characteristics of human interactions.
As a reminder, we only consider a potential interaction if both parties have WiFi scans within 300 seconds from one another.
For simplicity, we assume that the timestamp of the potential interaction is the lower of the two scan timestamps.
We notice that the prior probability of two people being proximate depends on the time of day and the day of week, as shown in Figure~\ref{fig:bt_wifi_features}I-K.
While there is only a small variability between the days of the week (Figure~\ref{fig:bt_wifi_features}J), the probability of the interaction during a day (Figure~\ref{fig:bt_wifi_features}I) appears to be driven both by the class schedule---the probability is the highest during classes, and drops during lunchtime---and by after-school social activities. 
Only by combining the two factors (Figure~\ref{fig:bt_wifi_features}K), we get the full picture: 
the probability of interactions from Monday to Tuesday is driven by the school schedule; 
Friday is a mixture of scheduled and social interactions, with the probability remaining high far into the night hours; 
Saturday is characterized by interactions starting in the late afternoon and into the night; 
and on Sunday our participants interact mostly during daytime, with no visible lunch breaks.
We add a feature to capture these patterns: 
\textbf{hour of week}: from 0 to 167.

\paragraph{Location. }
The last category, location, contains two binary features. 
A meeting is considered \textbf{at home} if at least one of the routers in the union corresponds to the home router of one of the users (the heuristic for home location detection is explained in~\ref{section:methods}).
A meeting is assumed to take place \textbf{at DTU} if at least one of the routers in the union broadcasts a WiFi network name of {\tt dtu}, as all access points on the campus do.




\subsection{Imputing missing values}
\label{section:imputing}
Two of our features are Pearson and Spearman correlations.
There are two cases in which it is not possible to calculate the correlation: (1) if there are fewer than three routers available for comparison, (2) if at least one person reads all the signal strengths at the same level.
In such cases we assume a NaN (not-a-number) value of $\rho$ to be imputed later on.
Additionally, we assume a NaN value of $\rho$ if the correlation is not significant with the $p_{val}<0.05$.
This results in multiple missing values for the two features.
The simplest approach is to skip such observations, but that would imply not training the model in cases with few routers available.
We therefore impute the values by assigning the mean value of the feature (averaged over all the non-NaN training examples) when we encounter NaN values.
This average from training is preserved and used to impute missing values in the test set.
We verified in our data that other approaches, such as using the median value of the feature or using $k$ nearest neighbors to impute the missing value~\cite{troyanskaya2001missing}, do not improve the consecutive predictive performance.



\section{Results}
\label{section:results}
In this section we evaluate the performance of each feature and each featureset in the task of proximity inference.
Then, we examine the robustness of our best model to short training as well as the various types of environments in which the interactions happen.

\subsection{Performance of single features}
We first show how well one can infer close-proximity interactions using single features.
We report the area under Receiver Operating Characteristic curve (AUC ROC) as the first metric of performance in Table~\ref{tab:1}.
Then, we select the threshold at which the $F_1$ score (the harmonic mean between precision and recall) is maximized in the training set.
We also report the $F_1$ score at the threshold optimal for the training set along with the AUC ROC for the test data (111.5 million previously unseen samples).

The results are presented in Table~\ref{tab:1}. 
We find that the single best performing feature is Jaccard similarity between the two lists of routers. 
As expected, thresholding on time information is not meaningful (it is equivalent to assuming that all interactions after a certain hour of a certain day of week are close proximity interactions).
It is important to note that the performance in test does not drop compared to training, which means that the thresholds are not just specific to the training data.

\subsection{Performance of feature sets}
We train a Gradient Boosting Classifier for each category of features and present the results in Table~\ref{tab:2}.
The parameters of the classifier are tuned each time through a grid search of the parameter space with 5-fold cross validation.
Furthermore, we compare the model based on the features proposed by Krumm \etal~\cite{krumm2004nearme} to models based on richer sets of features, see Table~\ref{tab:2}.
In the original work, Krum \etal~did not find any performance improvements of using a combined model over using single features.
Here, we show that combining the features they proposed does improve the performance.
Our Simple model is based on features that do not require long term data collection and are not specific to our deployment.
It performs better than any single feature or group of features, and it outperforms the model based on the features introduced by Krumm.
Enhancing the model with the information on popularity (the General model) further improves the performance.
Finally, using all features, including timing and location (which might be specific to this experiment as they depend on our campus as location and the time schedule typical for students), does not improve the performance of the classifier.


\begin{table}[t]
  \centering
  \begin{tabular}{r|l|c|c|c|c|}
 	\multicolumn{2}{c|}{} & \multicolumn{2}{c|}{AUC ROC} & \multicolumn{2}{c|}{$F_1$}\\ \hline
	category & feature & {train} & {test} & {train} & {test} \\ \hline
	\multirow{4}{*}{\makecell[r]{AP\\presence}} & overlap   & 0.77 & 0.77 & 0.61 & 0.61 \\ 
	& jaccard   & 0.84  & 0.84 & 0.69 & 0.68 \\
	& union  & 0.53 & 0.53 & 0.48 & 0.48 \\ 
	& non-overlap & 0.74 & 0.74 & 0.58 & 0.57 \\ \hline
	\multirow{4}{*}{RSSI} & spearman  & 0.70 & 0.70 & 0.57 & 0.58\\
	& pearson   & 0.71  & 0.71 & 0.59 & 0.59 \\
	& manhattan   & 0.60  & 0.60 & 0.51 & 0.51 \\ 
	& euclidean    & 0.59 &  0.59 & 0.51 & 0.51 \\ \hline
	\multirow{2}{*}{\makecell[r]{Presence\\+ RSSI}} & top AP & 0.60 & 0.60 & 0.48 & 0.48\\
	& top AP$\pm6dB$  & 0.75  & 0.74 & 0.65 & 0.65 \\ \hline
	\multirow{3}{*}{Popularity} & min\_popularity & 0.54 & 0.54 & 0.48 & 0.48 \\
	& max\_popularity &  0.59 & 0.59 & 0.49 & 0.50 \\ 
	& adamic\_adar & 0.77 & 0.77 & 0.62 & 0.62 \\ \hline
	Timing & hour of week &  0.51  & 0.51 & 0.48 & 0.48 \\ \hline
	\multirow{2}{*}{Location} & at DTU & 0.61 & 0.61 & 0.51 & 0.51  \\
	& at home & 0.64 & 0.64 & 0.55 & 0.55 \\ \hline

  \end{tabular}
  \caption{Performance of single features and feature categories in the task of inferring close proximity interactions.
  Jaccard similarity between lists of routers seen by the two devices is the best performing single feature.
  $F_1$ are given for a threshold that maximizes $F_1$ in the training set.}

  
  \label{tab:1}
\end{table}

\begin{table}[t]
  \centering
  \begin{tabular}{l|c|c|c|c|}
 	& \multicolumn{2}{c|}{AUC ROC} & \multicolumn{2}{c|}{$F_1$} \\ \hline
	featureset & {train} & {test} & {train} & {test} \\ \hline
	\makecell[l]{\textbf{AP presence}: overlap, \\non-overlap, jaccard,\\union} & 0.85 & 0.85 & 0.69 & 0.69 \\ \hline
	\makecell[l]{\textbf{RSSI}: spearman, \\pearson, manhattan,\\euclidean} & 0.78 & 0.79 & 0.62 & 0.62 \\ \hline
	\makecell[l]{\textbf{Presence+RSSI}: top AP, \\top AP$\pm6dB$} & 0.75 & 0.75 & 0.65 & 0.65 \\ \hline
	\makecell[l]{\textbf{Popularity}: min, \\max, adamic\_adar} & 0.79 & 0.79 & 0.62 & 0.62 \\ \hline
	\makecell[l]{\textbf{Location}: at DTU, \\at home} & 0.65 & 0.65 & 0.55 & 0.55 \\ \hline\hline
	\makecell[l]{\textbf{NearMe}: overlap, \\non-overlap, spearman,\\euclidean} & 0.87 & 0.87 & 0.71 & 0.71 \\ \hline
	\makecell[l]{\textbf{Simple}: \textbf{AP presence},\\\textbf{RSSI}, \textbf{Presence + RSSI}} & 0.88 & 0.88 & 0.72 & 0.72  \\ \hline
	\makecell[l]{\textbf{General}: \textbf{AP presence},\\\textbf{RSSI}, \textbf{Presence + RSSI}\\\textbf{Popularity}, at home} & 0.89 & 0.89 & 0.73 & 0.73 \\ \hline
	\textbf{Full}: all features & 0.89 & 0.89 & 0.73 & 0.73 \\ \hline 
  \end{tabular}
  \caption{Performance of feature sets in the task of inferring close proximity interactions. 
  We train a Gradient Boosted Classifier on selected subsets of features: 
  each feature category listed in Table~\ref{tab:1},
  NearMe~\cite{krumm2004nearme},
  Simple (no features that are specific to this experiment or require longer term data collection), 
  General (without features that could be specific to this experiment), 
  and Full (all listed features). 
  Using features which could be specific to the experiment does not improve performance further.}
  \label{tab:2}
\end{table}


\subsection{WiFi similarity and physical proximity}
\begin{figure}[t!]
\center
\includegraphics[width=1\linewidth]{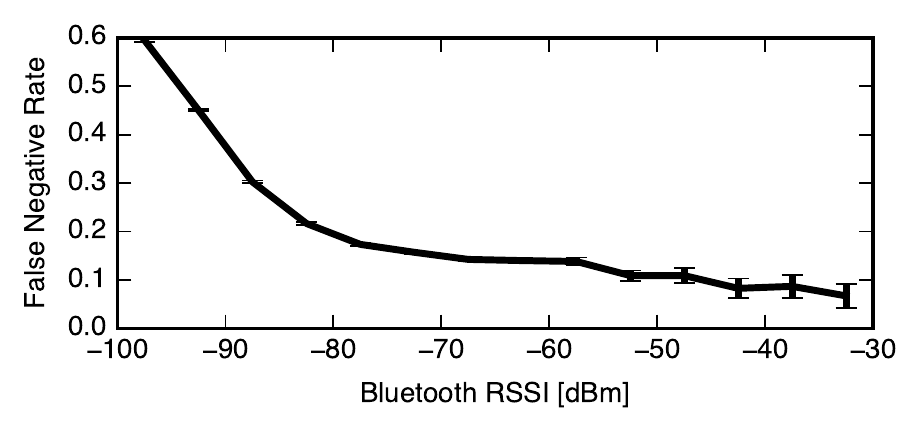}
\caption{The distance over which an interaction happens can be approximated using Bluetooth received signal strength (RSSI). Very close proximity contacts are unlikely to be misclassified as non-interactions. The lower the RSSI (the more distant the two potentially interacting people), the higher the probability, that our models miss the interaction.}
\label{fig:error_rssi}
\end{figure}

Here, we verify whether there is a correlation between how close people are in physical space (approximated by the received Bluetooth signal strength measured on their phones) and the probability that our models misclassify the sample as ``non-interaction''.
As we show in Figure~\ref{fig:error_rssi}, the shorter the distance over which an interaction happens (high Bluetooth RSSI), the lower the probability of missing that interaction.
This shows that the similarity measure between WiFi lists introduced by our models has a physical interpretation: a more similar WiFi environment indicates proximity in a more granular way than just the Bluetooth 10 meter range.

\subsection{Training period and performance in test}
\begin{figure}[t!]
\center
\includegraphics[width=1\linewidth]{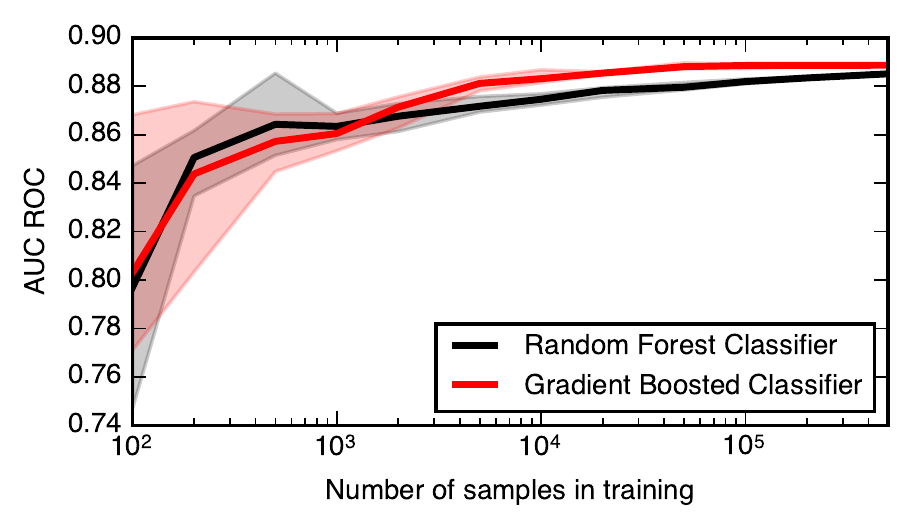}
\caption{The more samples we use for training the interaction detection models, the better they perform in test, but after a certain thresholds, the gains are negligible. The performance of the Gradient Boosted Classifier saturates at a higher level, but the time it takes to train the classifier is longer than it is the case with the Random Forest Classifier. Each of the model is trained 20 times for each number of samples, the shaded areas correspond to 25-75 percentiles and the solid lines to medians of the results for each training set size. }
\label{fig:training_period}
\end{figure}

Figure~\ref{fig:training_period} shows how the number of samples used for training influences the performance of the full model in test.
We compare the performance of a random forest classifier and a gradient boosted classifier and find that the latter has a slightly higher performance for training sets larger than 1000 samples.
On the other hand, training of the random forest classifier can parallelized, thus making the process faster.

\subsection{Importance of features}
\begin{figure}[t!]
\center
\includegraphics[width=1\linewidth]{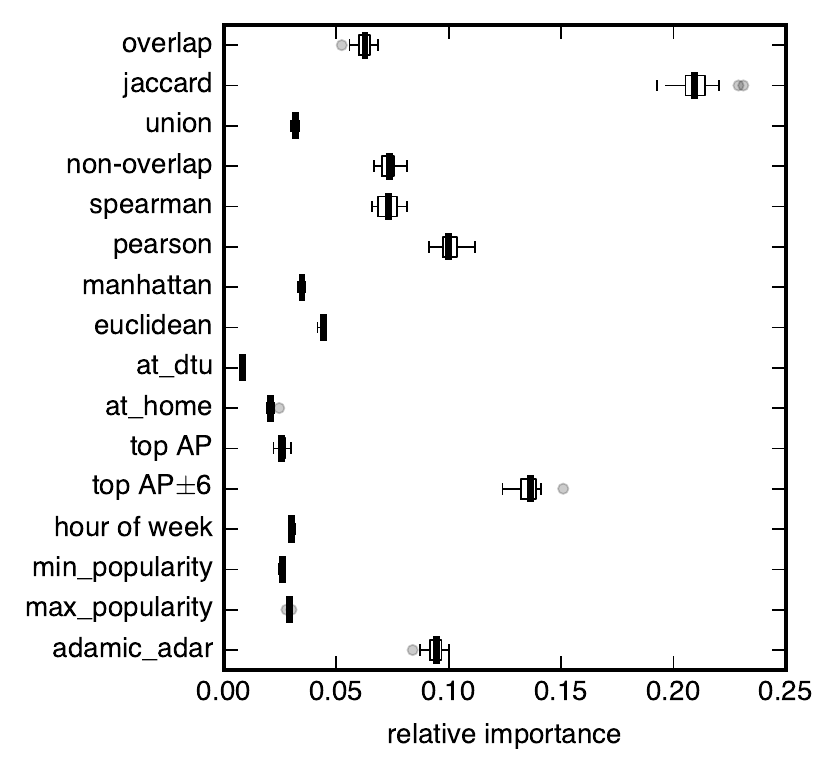}
\caption{Gradient Boosted Classifier reports the relative importance of each feature (the decrease in node impurity it provides). After 30 training rounds we see that Jaccard is the most important feature, followed by overlap among the strongest routers (top AP$\pm6dB$), Adamic-Adar, and Pearson correlation between the signal strengths. }
\label{fig:importances}
\end{figure}

Here we show how important each feature is for the machine learning model.
In the implementation we use~\cite{scikit-learn} the feature importance is defined as the total decrease in node impurity weighted by the probability of reaching that node, averaged over all trees of the ensemble~\cite{importance}.
Figure~\ref{fig:importances} shows the accumulated results from 30 training rounds of the gradient boosted classifier on randomly selected subsets of the training data, each with 100\,000 samples.
We find that Jaccard similarity is the most important, followed by the overlap among the strongest routers, Pearson's correlation of signal strengths, and Adamic-Adar (which exploits the overlap and the popularity of routers).

\subsection{Validity of the model in different scenarios}
\begin{figure}[t!]
\center
\includegraphics[width=1\linewidth]{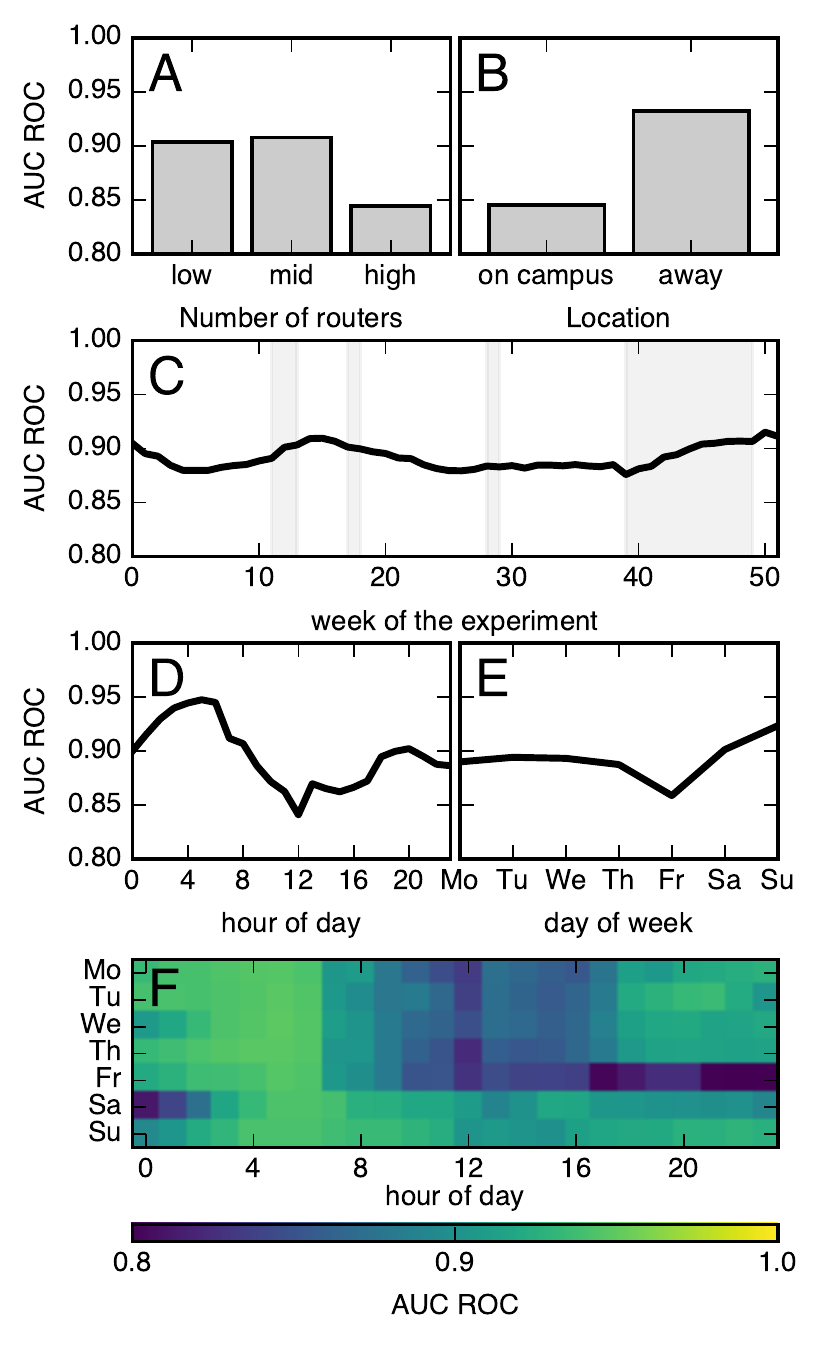}
\caption{Our model for detecting person-to-person proximity events performs well regardless of the number of available routers (A) and location (B). Its performance does not drop during holidays (marked with gray areas in C). The situation in which the performance is the worst is the Friday evenings and nights (F), but even then, the AUC ROC is high. }
\label{fig:all_results}
\end{figure}

Figure~\ref{fig:all_results} shows the performance of the gradient boosting classifier in different contexts and across time.

\paragraph{Number of routers. } 
As described before, the number of routers in an environment is positively correlated with the population density. 
We divide the test data in three equally-sized subsets, depending on the size of the union of routers seen by two people.
Figure~\ref{fig:all_results}A shows that the performance of the model is best in the low and mid sets ($AUC>0.9$) and observably lower ($AUC\approx0.85$) for environments with the highest number of routers.
Thus, we show that the model performs well in typical environments.

\paragraph{Location. }
Because our the data was collected by students of one university, with the majority of interactions happening on campus, there is a risk that the model would overfit towards such situation.
This is, in fact not the case.
Figure~\ref{fig:all_results}B shows that while the performance of the model is high on campus, it becomes even better for the meetings outside.

\paragraph{Timing. }
As shown in Figure~\ref{fig:all_results}C the performance of the model does not drop significantly during special periods, such as Christmas of summer vacation (gray areas in the plot correspond to periods with no university classes). Instead, it remains stable throughout the experiment.

The performance does vary with the hour of week, as shown in Figure~\ref{fig:all_results}D-F.
When we compare it to Figure~\ref{fig:bt_wifi_features}K, we see that the model performs better in situations where the prior probability of meeting is lower (for example during week nights).
Nevertheless, it retains high performance of $AUC\geqslant0.8$ throughout the week.

\section{Related work}

In this section we discuss related work that explores the application of mobile data to deepen our understanding of aspects relevant to this paper.

\paragraph{Location and mobility. }
CDR data has been used as a proxy for human mobility at large, societal scale.
It has been shown that our movements are regular~\cite{gonzalez2008understanding}, stable~\cite{Lu18062012}, and predictable~\cite{song2010limits}.
Several works argue that many unpredictable travels observed in real data can be attributed to individuals seeking interaction with their social contacts~\cite{grabowicz2014entangling,toole2015coupling,cho2011friendship}.
It yet remains to be verified whether these findings hold fully if the analysis were to be performed on data with higher spatial and temporal resolution (such as WiFi data).
At smaller scales, the scientific community investigated the potential of WiFi routers in applications of indoor~\cite{bahl2000radar,haeberlen2004,priyantha2000cricket} and outdoor~\cite{Cheng2005,skyhookWardriving,googleWardriving,han2009access} localization. 
Our recent work investigates how large companies can crowd source the creation of databases with router locations~\cite{sapiezynski2015opportunities,skyhookWardriving,googleWardriving} and how people's mobility on societal scale can be described using only a small subset of available routers~\cite{10.1371/journal.pone.0130824}.
WiFi signals can also be analyzed to discover places of interest and stop locations in an unsupervised manner, \textit{i.e.} without explicit location information as reference~\cite{kang2004extracting,wind2016inferring}.

It is important to stress that the work presented in this article does not rely on location estimation (in terms of geographical coordinates) but instead on relative comparison between the environments sensed by two parties.

\paragraph{Interactions. }
Complementary to mobility, the question of social interactions has been recently considered in various contexts, with the results indicating that collection of high-resolution behavioral traces is instrumental for understanding of complex processes in society~\cite{eagle2006reality, sekara2015fundamental, stopczynski2015temporal, stopczynski2015physical}.
However, from a technical point of view, collection of such data remains a challenge.

The most popular methods for quantitative and scalable collection of close-proximity interactions include using specialized hardware (\textit{e.g.}, sociometric badges)~\cite{olguin2009sensible, salathe2010high} or Bluetooth enabled smartphones~\cite{eagle2006reality, aharony2011social, 10.1371/journal.pone.0095978}.
In case of badges, interactions are usually inferred using radio-frequency identification (RFID) transmissions or infrared.
This way, badges worn around participants' necks can usually sense not just proximity but also whether individuals are facing each other, resulting in recordings of face-to-face interactions.
Sensing performed using Bluetooth-enabled mobile phones is less granular.
The proximity can be detected in a binary fashion or further refined using the received signal strength as a proxy for distance~\cite{sekara2014strength}.
However, the orientation of the individuals can not be sensed.
The subjects' devices must remain in Bluetooth-discoverable state, which raises a number of security and privacy concerns, as described in the Introduction.
There has been some developments in substituting Bluetooth with WiFi, an approach in which one of the phones acts as a hotspot and is sensed by others~\cite{carreras2012comm2sense}.
In controlled test environments this approach appears to offer a distance estimation resolution of 0.5m~\cite{osmani2014analysis}, providing a better understanding of the nature of the contacts~\cite{hall1966hidden}.
However, the claim has not been tested in the wild and the method potentially introduces even more privacy and security problems than Bluetooth.

An alternative way of sensing interactions between two persons with smartphones relies on comparing the two devices' radio frequency perceptions of the environment.
If a similarity is above a certain threshold, the two devices are assumed to be in physical proximity.
The idea of comparing WiFi signals to measure proximity was initially explored more than a decade ago.
Initially, researchers relied on single-feature measures of similarity, such as Manhattan distance~\cite{meunier2004peer} or overlap~\cite{McNett2005}.
NearMe project~\cite{krumm2004nearme} introduced more features, such as rank correlation between the lists of overlapping routers sorted by signal strength, Euclidean distance, and the number of non-overlapping APs.
The authors explored combining the features into a regression model, but this approach did not outperform single features.
Moreover, their model would overfit for the rooms where it was trained and thus under-perform in previously unseen environments.
In fact, Kj{\ae}rgaard and Nurmi name differences in environments where the sensing takes place among the most important obstacles in using WiFi for social sensing~\cite{kjaergaard2012challenges}.
Carlotto \etal combine a number of previously suggested features using a Gaussian Mixture Model and claim that their model is not environment-dependent (performs equally well in both buildings where it was tested)~\cite{Carlotto:2008:PCM:1410012.1410023}. 

We note that the differences in environments can actually be used to increase the performance of the model.
We can exploit the characteristics of human interactions: from a technical standpoint, environments with a smaller number of routers offer lower accuracy of distance estimation; however, two people in an environment with fewer access points are more likely to be actually interacting (see Figure~\ref{fig:bt_wifi_features}).

\section{Discussion}
In this paper we evaluated the applicability of WiFi based social sensing.
The idea of exploiting WiFi signals for this purpose is not new.
However, to our best knowledge, researchers have not yet tested this approach in practice, over a long period, and in a large population that interacts in various environments.
The growing popularity of WiFi access points and the phones' inability to remain Bluetooth discoverable are two trends that make it feasible and important to begin using WiFi signals for social sensing.

\subsection{Privacy implications}
There are two main privacy implications of this work.

\textbf{First}, the ability to track face-to-face interactions using WiFi can help us move away from relying on Bluetooth.
By not requiring the participants' phones to remain Bluetooth discoverable we protect the privacy and security of the subjects.
While currently most phones advertise their presence and identity by scanning for WiFi, this problem is being addressed.
Both Android and iOS randomize the MAC address of the device every time it sends WiFi probe requests making it more difficult to identify the user.\footnote{The randomization can only happen when the device is not connected to any WiFi network. When it is, it announces its real MAC address in each probe request.}

\textbf{Second}, our results indicate a potential erosion of privacy of Android users.
As we have previously shown, WiFi can be efficiently used for high-resolution mobility tracking of entire populations~\cite{sapiezynski2015opportunities, 10.1371/journal.pone.0130824,wind2016inferring}.
Here we go a step further and infer who people interact with, not only where they are.
Thus, results of WiFi scans---collected by major manufacturers of mobile devices and available to majority of mobile application developers---constitute very sensitive datasets.
For example, a vast majority of the applications available in Google Play Store has access to WiFi information, including all the scan results requested by the system as often as every 15 seconds~\cite{10.1371/journal.pone.0130824}.
This problem is addressed since Android 6.0---in the latest versions of the system an application has to hold a location permission to listen to WiFi scan results.
However, the vast majority of handsets currently in use will not receive these crucial updates.
Thus, WiFi signals remain a major privacy risk for years to come.

\subsection{Limitations of the WiFi-based social inference}
While our approach to inference of social interactions using WiFi signals offers an important new method in computational social science, we want to recognize its limitations.
The inference in the approach presented here depends on the WiFi routers being present in the environment.
While today WiFi networks are omnipresent, especially in densely-populated areas~\cite{10.1371/journal.pone.0130824}, we find that in our longitudinal and diverse dataset approximately 5\% of the WiFi scans did not report any nearby networks, preventing inference of physical proximity.

In this study, all phones collecting data were of the same make and model.
When considering a broader application of the method, differences in WiFi hardware transmitters and firmware and software of the phones may result in less consistent scan data, making it more difficult to devise a robust model as the one presented here.

Furthermore, due to the lack of ground truth data, we cannot prove that our model accurately estimates the distance between users.
We show, that our model is more likely to recognize interactions with a higher Bluetooth RSSI, but this property does not trivially translate to distance estimation.

Finally, we should note that it is not our argument that the values of all model features for discovering particular interactions and reconstructing the overall social network are generally applicable to different populations.
Depending on the specific population and social context under consideration, the weights in the model might be different or even entirely new features might be useful.
Our results indicate, however, that physical proximity can be inferred in a feasible fashion using WiFi signals collected by smartphones, even in very densely-connected populations.

\section{Conclusion}
In this work we showed how WiFi scan results can reveal a great deal about our daily interactions with others and our social ties.
By using behavioral traces, placed in context through meta information and our basic understanding of the inner working of social systems, we can transform a noisy data source to a strong social signal.
Our findings have important privacy implications, especially given our previous work which shows that it is possible to use WiFi signals for tracking human mobility.
On the other hand, WiFi scans also constitute a great opportunity for companies with access to such data on a global scale, to contribute \textit{e.g.}, better epidemic models built on proximity data of billions of people.
Finally, we hope that this method of social sensing will substitute Bluetooth sensing in future Computational Social Science deployments.

\section*{Acknowledgements}
The authors would like to thank Andrea Cuttone for useful discussions as well as Urvashi Khandelwal and Jana Huisman for the important feedback.
In this work we used the implementations of machine learning models from the scikit-learn~\cite{scikit-learn} Python package.

\bibliographystyle{abbrv}
\bibliography{bibliography}
\end{document}